# Structure related optical properties of electron beam evaporated $ZrO_2$:10%$SiO_2$ thin films

S. Jena[*], R. B. Tokas, S. Thakur and N. K. Sahoo

Atomic & Molecular Physics Division, Bhabha Atomic Research Centre
Trombay, Mumbai 400 085, India

**Abstract:** $ZrO_2$:10%$SiO_2$ thinfilms have been deposited on fused silica substrate by reactive electron beam co-evaporation technique at different oxygen partial pressure. The structural analysis shows tetragonal phase with residual tensile stress in the films. The intensity of the tetragonal t(110) phase are found increasing with increasing oxygen pressure. The optical band gap is found increasing from 5.06 eV to 5.28 eV because of increasing crystalinity of monoclinic phase, while the film grain size remains almost constant with increase of oxygen pressure, concludes that the crystallite or grain size has no effect on the optical properties of the films. The dispersion of the refractive index is discussed in terms of single oscillator Wimple-DiDomenico model. The dispersion energy parameter $E_d$ better known as structural order parameter are found increasing with the intensity of t(110) phase. It is observed that films having higher value of order parameter show lower surface roughness which concludes that the local microstructure ordering can predominantly influence the grain morphology which in turn can lead to better surface for higher value of order parameter.

*shuvendujena9@gmail.com

# 1. Introduction:

Zirconium oxide ($ZrO_2$) is a highly useful thin film material having potential applications in protective and thermal barrier coatings, optical filters, high power lasers, insulators, storage capacitors in dynamic random access memories and other electro-optic devices because of its outstanding thermal, chemical and mechanical stability with interesting optical and dielectric properties [1-3]. $ZrO_2$ films are known to exist in monoclinic phase, orthorhombic phase, tetragonal phase, cubic phase as well as amorphous structure [4, 5]. In recent years, $ZrO_2$ is stabilized by adding a fair amount of different foreign oxide materials, which leads to several distinct advantages such as tunability in the refractive index, optical band gap, decrease in intrinsic stress, smaller optical scatter, decreased surface roughness, enhanced laser damage threshold etc [6]. It is observed that by admixing silica in $ZrO_2$, under certain compositional ratios the optical properties such as refractive index, band gap etc are dominantly influenced by the microstructure and morphological evolutions rather than the stoichiometry. For instance, by adding 10 to 20 % of $SiO_2$, the $ZrO_2$: X%$SiO_2$ (X=10-20) binary oxide films show higher refractive index as well as higher laser induced damage threshold with better surface morphology as compared to the pure $ZrO_2$ [7, 8]. But the performance of the multilayer coatings does not depend solely on its optical properties, but also on its mechanical properties such as residual stress and adhesion. Excessive residual stress can limit the reliability and function of thin film based structures due to peeling, cracking and curling [9]. Hence for the fabrication of high quality optical coatings, control over residual stress is very crucial. It is well known that deposition oxygen partial pressure is one of the important parameters to alter the properties of thin films. Therefore, the effect of oxygen pressure on the microstructural and optical properties of the $ZrO_2$:10% $SiO_2$ binary oxide films has been investigated in order to find suitable deposition conditions for better quality films. Grazing incidence x-ray diffraction (GIXRD) has been carried out in order to examine the crystallization behaviour as well as to determine the residual stress of the films. The optical properties, surface morphology and grain size of the films were studied using spectrophotometer, atomic force microscopy (AFM) and field emission scanning electron microscopy (FE-SEM) respectively. The films prepared at higher oxygen pressure are found to exhibit lower residual stress and higher refractive index. It is observed that the films having better surface shows higher value of microstructure order parameter $E_d$ which confirms that the morphology is strongly influenced by the local microstructure. Interestingly, no crystallite size or grain size effects on the optical properties for the reactive electron beam evaporated $ZrO_2$:10% $SiO_2$ films has been observed as reported by C.V. Ramana *etal* [10] for magnetron sputtered $ZrO_2$ films under varying growth temperature, as the film grain size is found almost same while the optical band gap increases with increasing oxygen pressure in the present study.

## 2. Experimental details

The $ZrO_2$:10%$SiO_2$ composite thin films at different oxygen partial pressure have been deposited on fused silica substrates by reactive electron beam co-deposition technique in a fully automatic thin films deposition system VERA-902. The film materials for $SiO_2$ and $ZrO_2$ were chosen from Cerac's batch number "S-1060" (purity 99.99%) and "Z-1056" (purity 99.995%) respectively. The system is equipped with two 8 KW electron beam guns with sweeps and automatic emission controls. The real time deposition rate and physical thickness were controlled and determined by Inficon's XTC/2 quartz crystal monitors (QCM), while the optical thickness of the composite films was measured by Leybold's OMS-2000 optical thickness monitor. The total pressure inside the vacuum system during reactive evaporation process was controlled using MKS mass flow controllers. The deposition rate 9Å/sec for zirconia and 1Å/sec for silica, and the substrate temperature of 300°C were maintained during deposition for all the $ZrO_2$:10%$SiO_2$ films while the oxygen pressure was varied from 0 mbar to $8\times10^{-4}$ mbar. All these films studied here are having optical thickness of 6 to 8 quarter wave thick at the monitoring wavelength of 600 nm in order to have appropriate numbers of interference fringes for spectrophotometric analysis. The structural characterization has been carried out by GIXRD using Bruker's D8 Advanced XRD unit with $2\theta$ angle in the range of $25^0$-$80^0$ using Cu-$K_\alpha$ ($\lambda$=1.5406 Å) radiation in steps of $0.05^0$. Since the measurements have been carried out on thin film samples the data were recorded by keeping the incident angle fixed at a grazing angle of incidence of $2^0$ in order to keep the probing region of the sample near its surface. The phase identification is achieved by comparison with data from the ICDD international diffraction database. The optical transmission spectra of the films were measured using UV-VIS-NIR spectrophotometer (UV-3101PC, SHIMADZU) in the range 190-1200 nm for determination of optical constants and thickness of the films. Atomic force microscopy (AFM) has been used to estimate the surface roughness using NT-MDT's Solver P47H multimode scanning probe system. AFM images were obtained in non-contact mode using silicon nitride cantilever. The grain size as well as grain distribution have been analyzed using a JEOL (JSM-7600F) Field Emission Scanning Electron Microscope (FEG-SEM).

### 3. Results and discussion

GIXRD patterns of $ZrO_2$:10%$SiO_2$ films prepared at different oxygen pressure are shown in Fig.1. Without any oxygen pressure (0 mbar $O_2$), the peaks are all attributed to diffraction from different planes of zirconia tetragonal phase. With increasing oxygen pressure, the film shows more crystallization behaviour. The peaks at approximately $35^0$ and $60^0$ are attributed to the diffraction from (110) and (211) planes of tetragonal phase of ZrO2. Diffraction peaks from the monoclinic phase is very weak, which indicates that fraction of monoclinic phase in the films is very low. The dominant phase in the film is the tetragonal. In all cases, the films show random orientation or polycrystalline nature. The intensity of the peak t(110) increases with increasing deposition oxygen pressure. At the same time, a new diffraction peak from m(111) appears at an oxygen pressure of $2\times10^{-4}$ mbar and the intensity of it increases with the increase of oxygen pressure. The x-ray diffraction data can be used to determine the residual stress in the films. The displacement of diffraction peaks from their corresponding stress-free data indicates a uniform stress developed normal to the corresponding crystal plane in the film during growth of the film [11]. This in-plane stresses may either be compressive or tensile which is associated with a contraction or an elongation of lattice spacing along in-plane directions. The most intense peak (110) of the XRD patterns has been used for the quantitative analysis of residual stress. The peak (110) shifts to higher $2\theta$ values with respect to the bulk $ZrO_2$ and is attributed to tensile stress in the films. The residual stress of the films has been calculated using the following equation [12]

$$\sigma = \left(-\frac{E}{\gamma}\right)\frac{(d-d_0)}{d_0} \quad (1)$$

Where d is the inter planer spacing of the films corresponding to (110) plane, $d_0$ is the standard plane spacing corresponding to the stress-free system (2.544 Å) of x-ray diffraction file PDF#881007, $E = 170$ GPa is the Young's modulus and $\gamma = 0.28$ is the Poisson ratio [12]. The values of $d$ are calculated using the formula $2d \sin\theta = \lambda$, where $\theta$ is the Bragg angle corresponding to (110) peak in the XRD pattern and $\lambda=1.5406$ Å is the wavelength of x-ray. The value of d is found to be varying from 2.529 Å to 2.541 Å for the films prepared at different oxygen partial pressure. The residual stress in the films is estimated using equation (1) and is given in Table-1. The positive sign of σ indicates that the stress in the films is tensile in nature. The total residual stress in a thin film is composed of intrinsic stress, thermal stress and extrinsic stress [13]. It is well known that the intrinsic stress strongly depends on deposition process, while the thermal stress is caused both by the thermal expansion co-efficient difference between the film $\alpha_f$ and the substrate $\alpha_s$ and by the temperature difference between the deposition temperature $T_d$ and the ambient temperature $T_0$. It can be calculated by $\sigma_{th} = \left(\frac{E}{\gamma}\right)(\alpha_f - \alpha_s)(T_d - T_0)$. Since $\alpha_f = 10.2 \times 10^{-6} K^{-1}$ for ZrO$_2$ is greater than that of the substrates $\alpha_s = 0.54 \times 10^{-6} K^{-1}$ for fused silica and the deposition temperature is 300 $^0$C, so the thermal stress of 684 MPa is tensile in nature. But the estimated residual stress is greater than that of thermal stress except for the film prepared at 4x10$^{-4}$ mbar oxygen pressure, which indicates that the thermal stress is only a small fraction of the residual stress. The film prepared at 4x10$^{-4}$ mbar oxygen pressure shows least residual stress and the value is lower than the estimated thermal stress, which concludes, residual stress can be minimized under certain deposition condition by minimizing the intrinsic stress.

The spectra transmittance of ZrO$_2$:10%SiO$_2$ films with different oxygen pressures are shown in Fig.2. The inset plot in this figure shows the transmittance spectra of the films near the absorption edge. The relatively lower transmittance at low oxygen pressures may be due to nonstoichiometric ZrO$_2$:10%SiO$_2$ composite. The improvement in the transmittance at higher oxygen pressure clearly indicates that the film stoichiometry gradually improved as the oxygen partial pressure increased. The optical constants such as refractive index and absorption co-efficient as well as thickness of the films are evaluated using inverse synthesis method [14, 15]. The dispersion model used for describing the refractive index spectra is based on the single effective oscillator model according to Wemple and DiDomenico [14] and is given by

$$n(\lambda) = \sqrt{1 + \left(\frac{E_0 E_d}{E_0^2 - (1240/\lambda)^2}\right)} \qquad (2)$$

Where $\lambda$ is the wavelength (in nm) of the light, $E_0$ is the single oscillator energy, and $E_d$ is the dispersion energy. The imaginary part of the refractive index, i.e., the extinction coefficient (k) of the films is assumed to follow the Urbach model [15] and is given by

$$k(\lambda) = \alpha \exp\left(1240\beta\left(\frac{1}{\lambda} - \frac{1}{\gamma}\right)\right) \quad (3)$$

The experimental transmission spectrum is fitted using the co-efficient of the dispersion model and the thickness of the film as fitting parameters, where the fitting has been carried out using the formalism described elsewhere [14, 15]. Finally these fitting parameters are used to determine the thickness ($t$) and the refractive index ($n$) and extinction coefficients ($k$) spectra of the films. The fitting parameters and the thickness values of the films obtained as above have been given in Table-1. Fig.3 shows the representative experimental transmission spectrum of the film prepared at oxygen pressure of $2 \times 10^{-4}$ mbar along with the best-fit theoretical spectrum. The refractive index for entire $ZrO_2$:10%$SiO_2$ samples is shown in the inset plot of Fig.3. The optical band gap of the films has been estimated from the obtained values of absorption co-efficient ($\alpha$). To that end, it should be considered that the absorption co-efficient of the films in the high absorption region is given according to Tauc's relation for the allowed indirect transition [16] by the following relation

$$(\alpha E)^{1/2} = K(E - E_g) \quad (4)$$

Where $K$ is constant depends on the transition probability and $E_g$ is the optical band gap. Fig.4 is a typical best fit of $(\alpha E)^{1/2}$ vs. photon energy $E$ for the films at different oxygen partial pressure. The values of the optical band gap $E_g$ were taken as the intercept of $(\alpha E)^{1/2}$ vs. $E$ at $(\alpha E)^{1/2} = 0$. The optical band gap derived for each film is given in Table-1. It clearly shows that the band gap depends on the oxygen partial pressure and varies from 5.06 eV to 5.28 eV with increase of oxygen pressure.

It is interesting to note that the films surface roughness is lower at higher oxygen pressure as given in Table-1. By comparing AFM image of the films in the Fig. 5, it can be seen that the film prepared at without oxygen pressure (0 mbar) is more void rich structure showing a root mean square (rms) surface roughness of 3.51 nm, while the film prepared at $8 \times 10^{-4}$ mbar shows more dense structure with surface roughness of 2.25 nm, indicative for a very smooth film. The SEM image of the films is shown in Fig. 6. It is observed that the average grain size of all the films prepared at different oxygen pressure is found almost same except for the film prepared at $0.6 \times 10^{-4}$ mbar. Fig. 7(a) and (b) shows the variation of refractive index and residual stress with the deposition oxygen pressure. It can be clearly noticed that the film with higher residual stress has least refractive index and vice versa. This behaviour of refractive index may be due to the variation in inter-planar distance of the nano-crystalline films as observed by the XRD pattern. The film deposited at 0 mbar oxygen pressure having least refractive index shows poor crystalline structure as compared to the film prepared at $8 \times 10^{-4}$ mbar. The interplanar distance strongly affects the film packing density. The film having larger interplanar distance between lattice planes compared to the stress-free system shows poor film packing density, hence poor refractive index. This suggests that the refractive index strongly depends on the microstructure of the film. In the present study, the band gap is found increasing with oxygen pressure as shown in Fig.8. Since the grain size of the films prepared at different oxygen partial pressure is found almost same within the error bar as shown in Fig.8, hence the possibility of quantum confinement effect has been discarded. But, from the structural analysis of the $ZrO_2$:10%$SiO_2$ films, the increase of band gap energy can be related to the enhanced monoclinic phase with the oxygen pressure. Since the monoclinic-$ZrO_2$ phase has higher optical band gap as compared to that of tetragonal phase, so the shape of fundamental optical absorption can be different for films that contain a monoclinic-$ZrO_2$ component compared to films that are single tetragonal-$ZrO_2$ phase [17, 18]. The enhanced monoclinic phase of $ZrO_2$ may result in blue shift of band gap energy. Through analysing more than 100 widely different solids and liquids, the dispersion energy parameter $E_d$ is found dependent on the co-ordination number and atomic valency uniquely [19]. The dispersion energy $E_d$ is an important parameter that measures the average strength of interband optical transitions and is associated with the changes in the structural order of the material, the more ordered the materials the larger the dispersion energy $E_d$ and that is the reason it is treated as a microstructural ordering parameter [20]. The values of the dispersion energy ($E_d$) as a function of x-ray diffraction intensity of t(110) along with a parabola fitting curve are presented in Fig.9. It shows that $E_d$ at the largest intensity of t(110) has the largest value i.e. the structural order of the films prepared at $8 \times 10^{-4}$ mbar oxygen pressure is the highest which corresponds with the structural change of the film seen from the XRD patterns. This type of correlation between the dispersion energy or order parameter $E_d$ and the intensity of XRD pattern has also been observed in the $ZrO_2$ films prepared by RF sputtering at different substrate bias [21]. From Fig.10, it is clear that there exists a correlation between the order parameter $E_d$ and the rms surface roughness, the film having higher $E_d$ has the least rms surface roughness. Hence, the increase of the dispersion energy $E_d$ is associated with evolution of the thin film microstructure to a more ordered phase and better grain morphology as confirmed by both the GIXRD and AFM as shown in Fig.9 and Fig.10 respectively.

**Conclusion:**

The microstructure dependent optical properties of the $ZrO_2:10\%SiO_2$ thin films have been investigated. It is found that all the films exhibits strong tetragonal phase t(110) while the films prepared at $2\times10^{-4}$ mbar oxygen pressure or more also contain weak monoclinic phase m(111) and the crystalinity increases with increasing oxygen pressure. The film with higher residual stress has found to show least refractive index and vice versa. This behaviour of refractive index has been ascribed to the variation in inter-planar distance of the nano-crystalline films. Optical band gap of the film is found to be affected by the monoclinic crystalinity phase. No size effect on the optical properties has been observed. The dispersion energy or order parameter $E_d$ of the films has the largest value at the largest intensity of t(110) which correspond with the structural order of films seen from the XRD pattern. A correlation between the order parameter $E_d$ and the rms surface roughness is observed. The film having higher $E_d$ has the least rms surface roughness which corresponds to better grain morphology seen from the AFM image. The results of the study conclude that the optical properties of the films strongly depend on the microstructure influenced by the deposition oxygen pressure.

# Captions to Figures

**Fig. 1:** GIXRD pattern of $ZrO_2$:10%$SiO_2$ thin films prepared at different oxygen partial pressures.

**Fig. 2:** Transmission spectra of $ZrO_2$:10%$SiO_2$ thin films prepared at different oxygen partial pressure.

**Fig. 3:** Experimental transmission with best fit theoretical simulation of $ZrO_2$:10%$SiO_2$ thin film prepared at $2 \times 10^{-4}$ mbar oxygen pressure and refractive index spectra of all the thin films (inset plot).

**Fig. 4:** Plot of $(\alpha h\nu)^{1/2}$ vs. $h\nu$ for all the $ZrO_2$:10%$SiO_2$ thinfilms.

**Fig. 5:** 3-D AFM surface morphology of the thin films prepared at different oxygen partial pressure (a) 0 mbar (b) $0.6 \times 10^{-4}$ mbar (c) $4 \times 10^{-4}$ mbar (d) $8 \times 10^{-4}$ mbar.

**Fig. 6:** FESEM image of the thin films prepared at different oxygen partial pressure (a) 0 mbar (b) $0.6 \times 10^{-4}$ mbar (c) $2 \times 10^{-4}$ mbar (d) $8 \times 10^{-4}$ mbar.

**Fig. 7:** Variation of (a) refractive index and (b) residual stress with deposition oxygen partial pressure.

**Fig. 8:** Variation of band gap and grain size with deposition oxygen partial pressure and Plot of band gap vs. average grain size for all the films (inset plot).

**Fig. 9:** The order parameter $E_d$ as a function of intensity of t(1 1 0) from the XRD pattern of the films deposited at different oxygen partial pressures.

**Fig. 10:** The surface roughness as a function of order parameter $E_d$ for the films prepared at different oxygen partial pressure.

# Caption to table

1. Table-1: Parameters derived from GIXRD, transmission, AFM and FESEM measurement of all the $ZrO_2$:10%$SiO_2$ films

**Fig. 1**

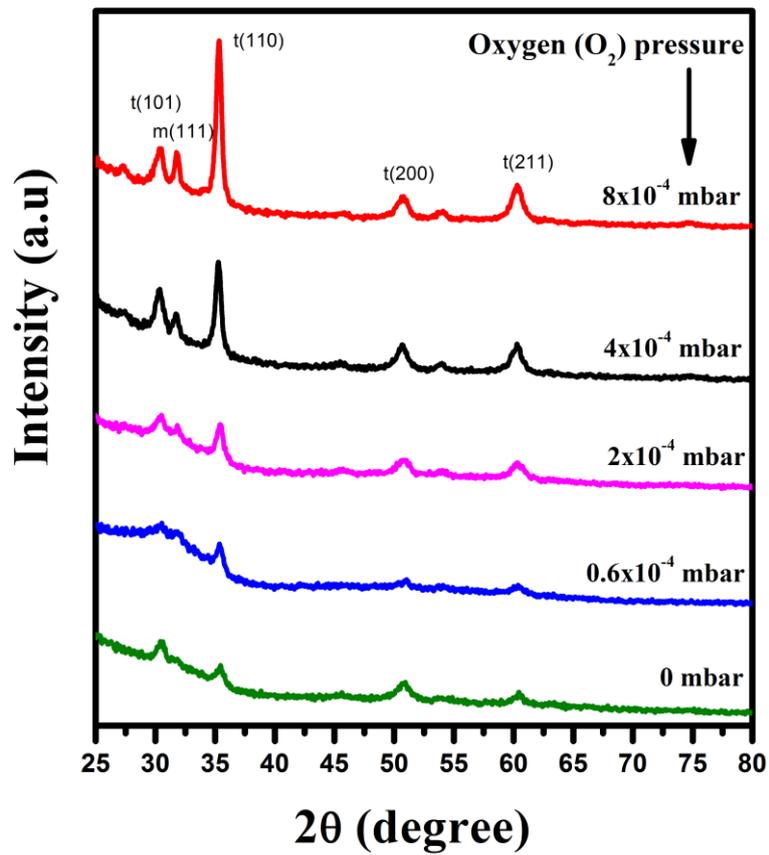

**Fig. 2**

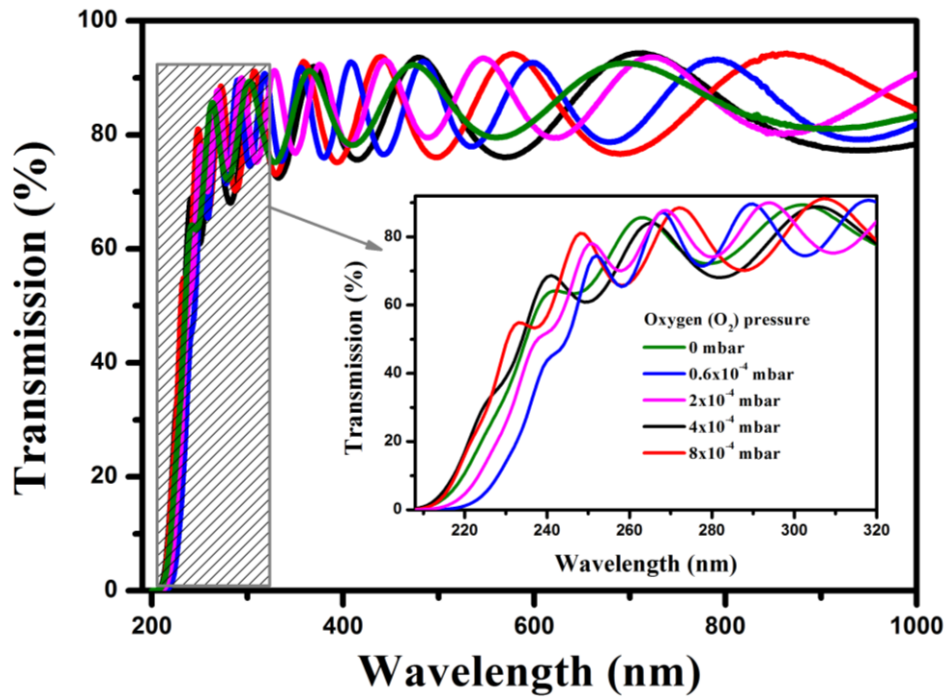

**Fig. 3**

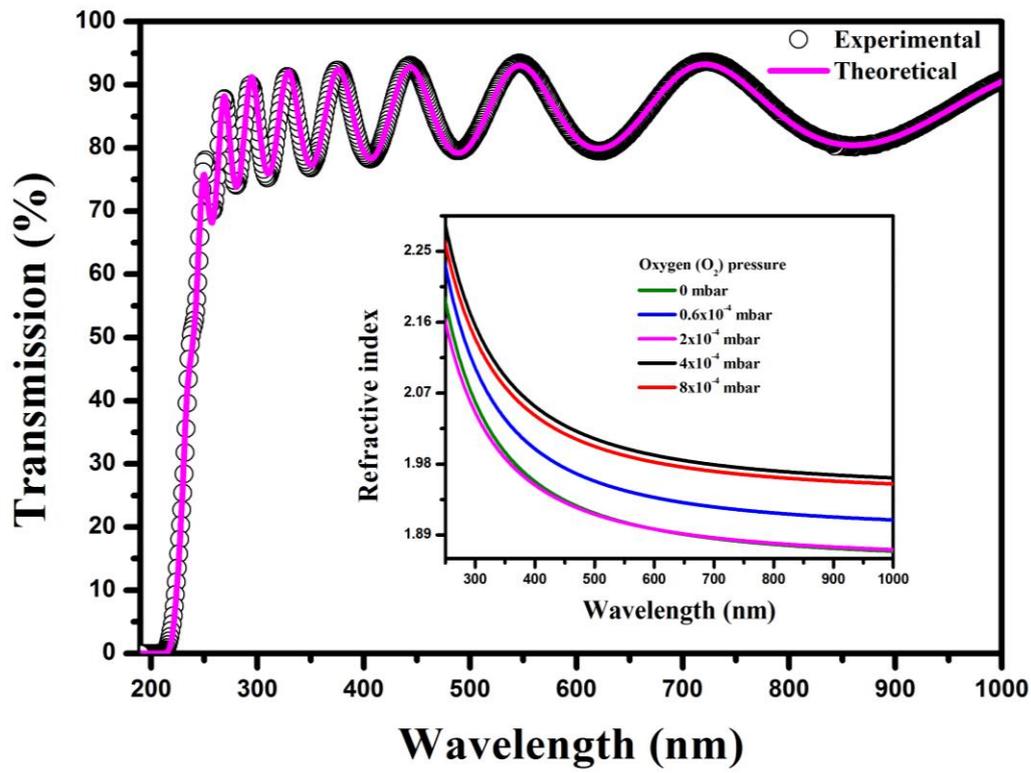

**Fig. 4**

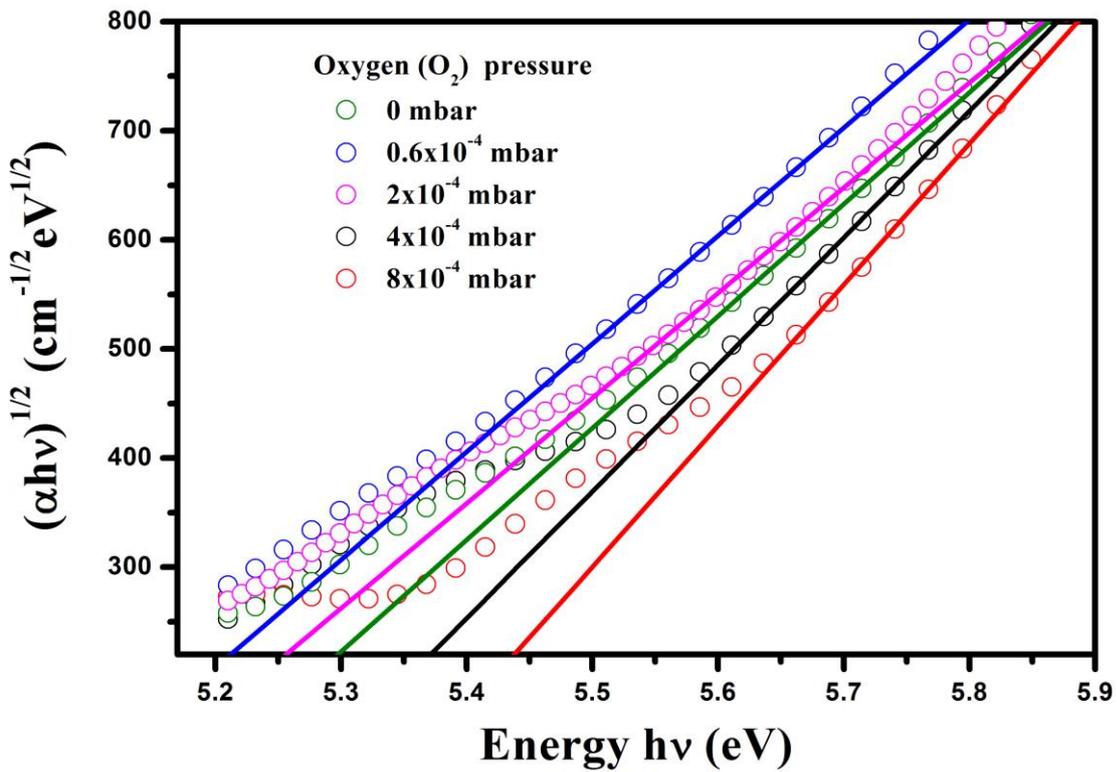

**Fig. 5**

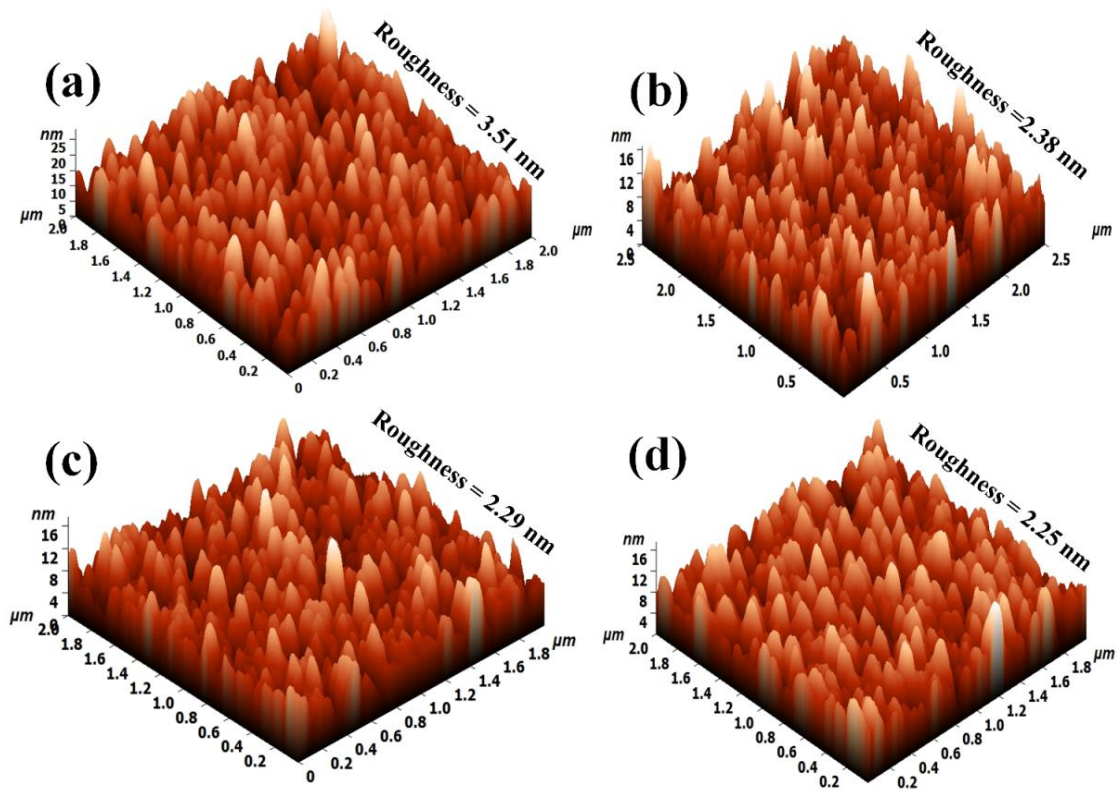

**Fig. 6**

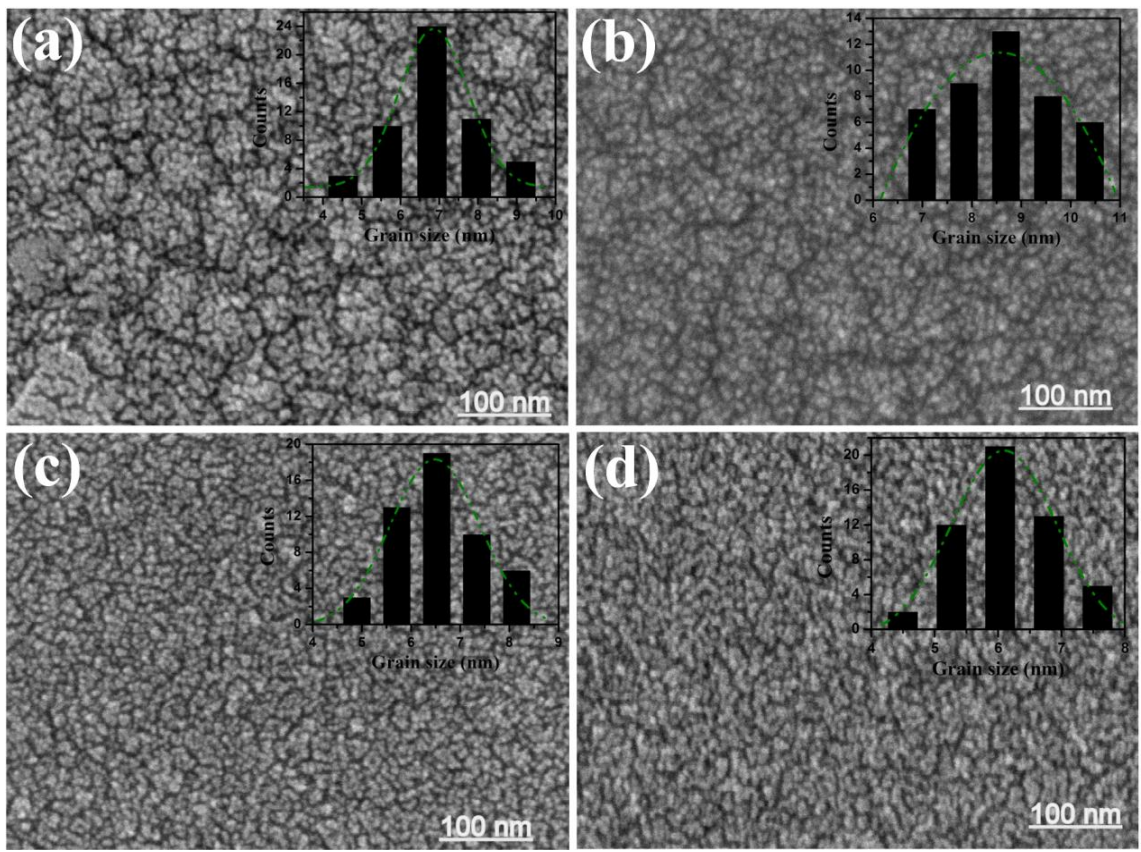

**Fig. 7**

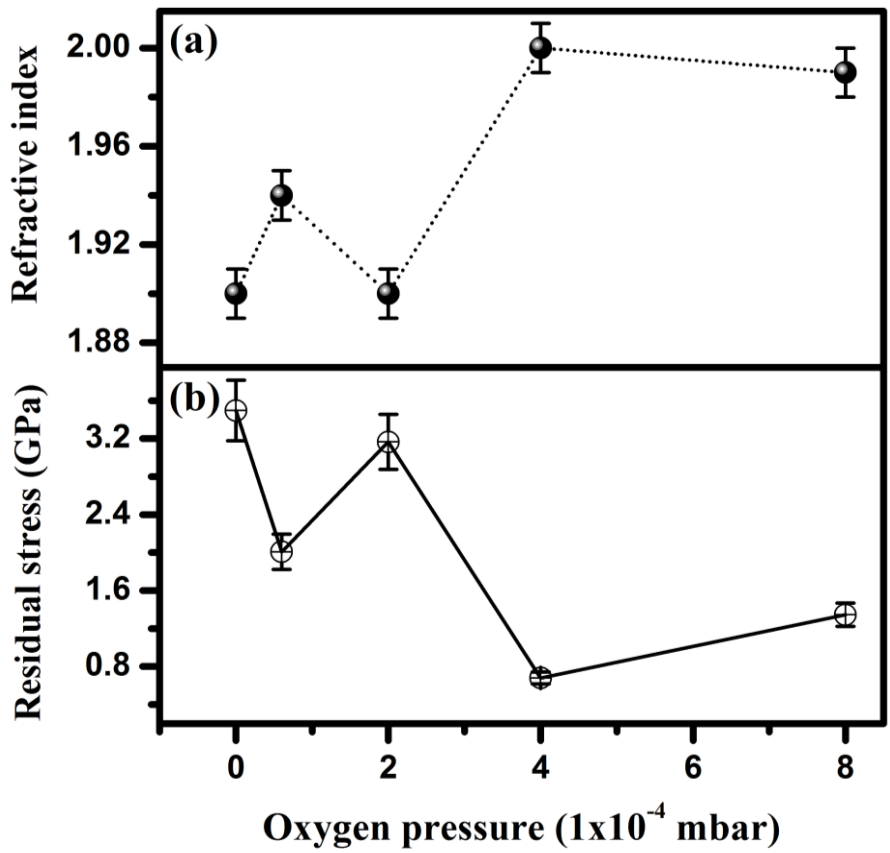

**Fig. 8**

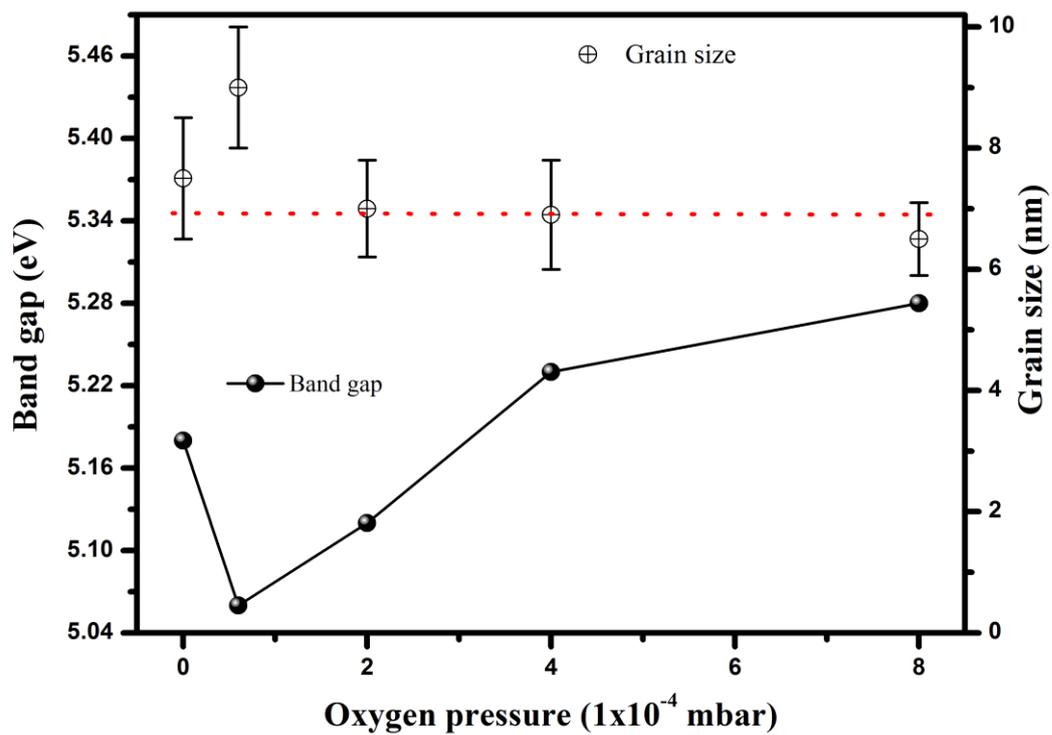

**Fig. 9**

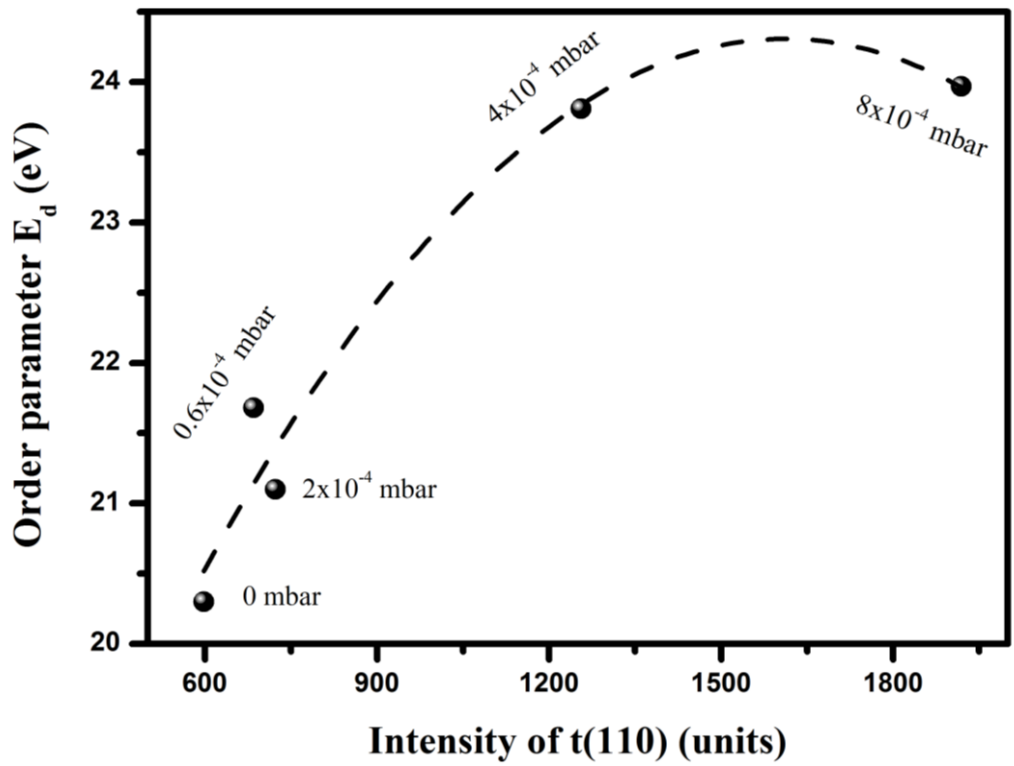

**Fig. 10**

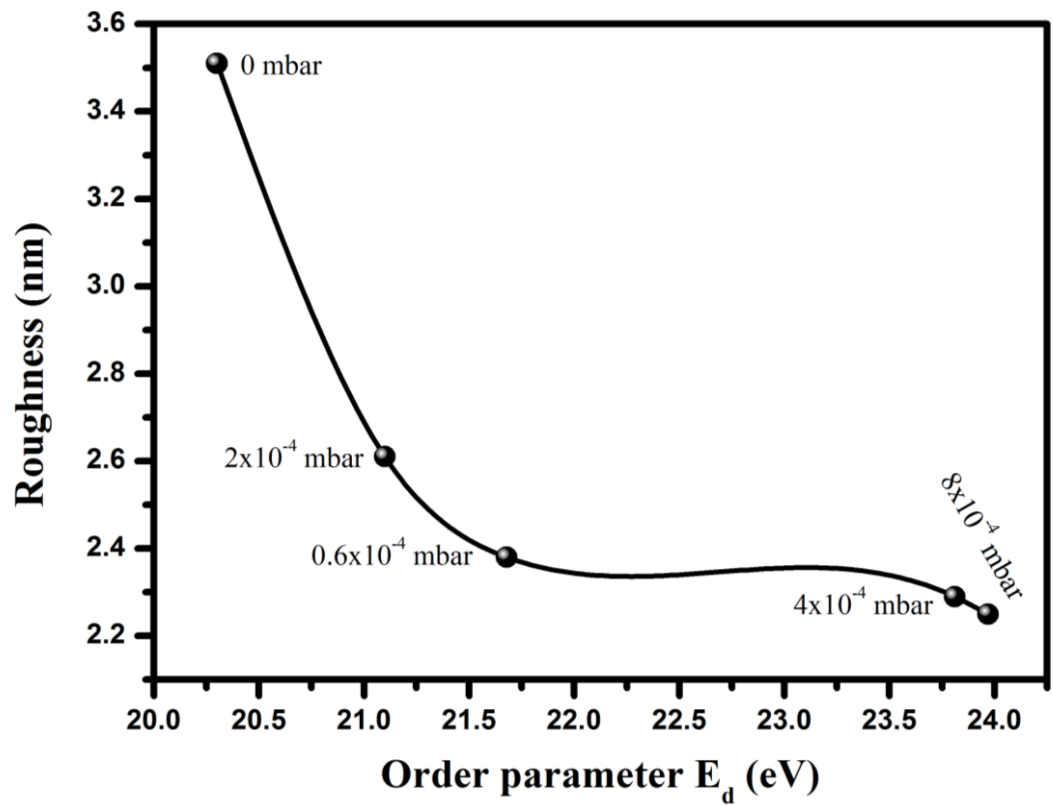

**Table-1**

| Oxygen pressure (mbar) | $2\theta$ (degree) | $d_{110}$ (Å) | Residual stress (GPa) | $E_d$ (eV) | $E_0$ (eV) | Thickness derived from transmission (nm) | Thickness in-situ measured (nm) | $E_g$ (eV) | Roughness (nm) | Grain size (nm) |
|---|---|---|---|---|---|---|---|---|---|---|
| 0 | 35.45 | 2.529 | 3.49 | 20.3 | 8.32 | 368 | 333±17 | 5.18 | 3.51 | 7.5±1 |
| 0.6 × 10$^{-4}$ | 35.37 | 2.535 | 2.00 | 21.68 | 8.38 | 615 | 651±32 | 5.06 | 2.38 | 9±1 |
| 2 × 10$^{-4}$ | 35.44 | 2.530 | 3.16 | 21.10 | 8.61 | 573 | 493±25 | 5.12 | 2.61 | 7±0.8 |
| 4 × 10$^{-4}$ | 35.29 | 2.541 | 0.67 | 23.81 | 8.53 | 356 | 343±18 | 5.23 | 2.29 | 6.9±0.9 |
| 8 × 10$^{-4}$ | 35.33 | 2.538 | 1.34 | 23.97 | 8.67 | 435 | 391±20 | 5.28 | 2.25 | 6.7±1 |